# Network Selection Using TOPSIS in Vertical Handover Decision Schemes for Heterogeneous Wireless Networks

**K.Savitha[1], DR.C.Chandrasekar[2]**

**[1] Research Scholar, Periyar University**
**Salem, Tamil Nadu, India.**

**[2] Associate Professor, Department of Computer Science, Periyar University**
**Salem, Tamil Nadu, India.**

**Abstract**

"Handover" is one of the techniques used to achieve the service continuity in Fourth generation wireless networks (FGWNs). Seamless continuity is the main goal in fourth generation Wireless networks (FGWNs), when a mobile terminal (MT) is in overlapping area for service continuity Handover mechanism are mainly used While moving in the heterogeneous wireless networks continual connection is the main challenge. Vertical handover is used as a technique to minimize the processing delay in heterogeneous wireless networks this paper, Vertical handover decision schemes are compared and Technique of order preference by similarity to ideal solution (TOPSIS) in a distributed manner. TOPSIS is used to choose the best network from the available Visitor networks (VTs) for the continuous connection by the mobile terminal. In our work we mainly concentrated to the handover decision Phase and to reduce the processing delay in the period of handover

## 1. Introduction

One of the main goal and most challenged area in fourth generation wireless network (FGWN) was service continuity, i.e., when a mobile node is moving in an overlapping area continuous service need so the handover technique is used. While a mobile terminal (MT) is moving from one network to another network is called heterogeneous network which has different air interfaces techniques there a handover techniques is used called Vertical Handover (VHO) .VHO is mainly used to support between different air interfaces techniques during inter-network movements.

"HANDOVER" is used to redirect the mobile user service from current network to a new network; handover is mainly used to select the suitable network to connect after the handover execution with minimum processing delay. In Fig. 1 Handover (HO) mechanism is illustrated

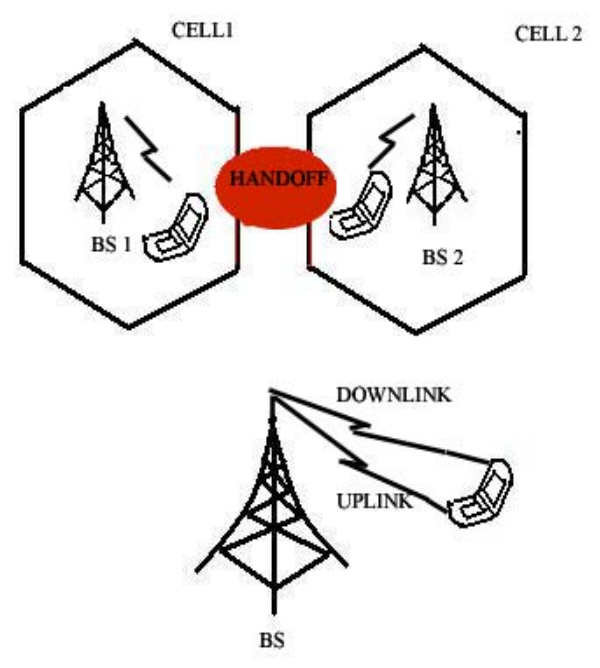

Fig. 1 Handover mechanism

In handover there are two types, one is Horizontal Handover (HHO) and another is Vertical Handover (VHO) as in Fig. 2. Horizontal handover is the process when the mobile user switching between the network with the same technology and same networks like Wifi to WiFi.

Vertical handover is the process when the mobile users switching among the networks with different technologies like WiFI to WiMax. So in heterogeneous networks VHO is mainly used.

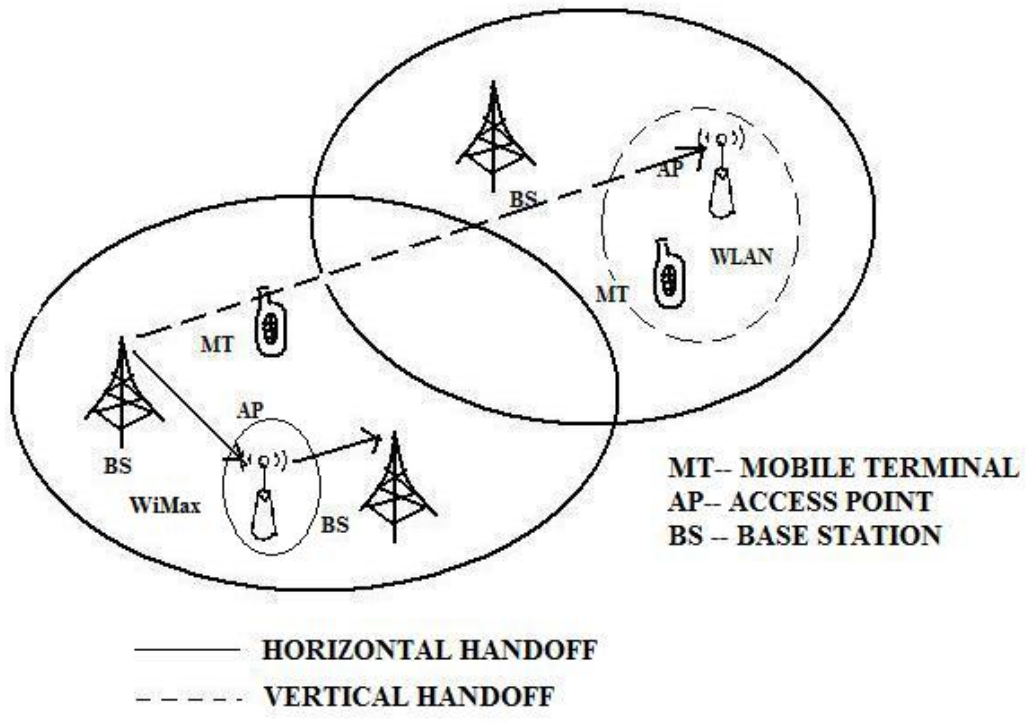

Fig. 2 Horizontal vs Vertiacl Handover

Handover mechanisms have a four different phases: Handover Initiation, System discovery, Handover decision, Handover execution.

- Han doff Initiation phase: The handover process was modified by some criteria value like signal strength, link quality etc.,
- System discovery phase: It is used to decide which mobile user discovers its neighbour network and exchanges information about Quality of Service (QOS) offered by these networks.
- Handover Decision phase: This phase compares the neighbour network QOS and the mobile users QOS with this QOS decision maker makes the decision to which network the mobile user has to direct the connection.
- Handover Execution phase: This phase is responsible for establishing the connection and release the connections and as well as the invocation of security service.

In our work HO decision phase is mainly focused for decision maker to choose a best network from a set of available visited network (VN). Multiple attribute decision making (MADM) - TOPSIS is used as decision maker to choose the best network and redirects the connection to the MT.

MADM is mainly concentrates on predetermined set of alternatives, i.e. the selection of an alternative from a discrete decision space and the methods are used when finite number of alternatives with associated information on regarded criteria is given. MADM have several cardinal information methods like simple additive weighting (SAW), weighted product method, Analytic Hierarchy Process (AHP), TOPSIS etc.,

In this paper, two vertical handover decision schemes (VHDS) , Distributed handover decision scheme (DVHD) and Trusted Distributed vertical handover decision schemes (T-DVHD)are used. DVHD is advanced than the centralised vertical handover decision scheme and T-DVHD is the extended work of DVHD. Here we compare the distributed and trusted vertical handover decision schemes as distributed decision tasks among networks to decrease the processing delay caused by exchanging information messages between mobile terminal and neighbour networks. To distribute the decision task, vertical handover decision is formulated as MADM problem.

The scope of our work is mainly in handover decision phase, as mentioned in the decision phase; decision makers must choose the best network from available networks. In this paper, the decision maker Technique for order preference by similarity to ideal solution to take the decision and to select the best target visitor network (TVN) from several visitors' networks.

The proposed decision making method use TOPSIS in a distributed manner. The bandwidth, delay, jitter and cost are the parameters took by the MT as the decision parameters for handover.

## 2. Related Work

Many of the handover decision algorithms are proposed in the literature In [1] a comparison done among SAW, Technique for Order Preference by Similarity to Ideal Solution (TOPSIS), Grey Relational Analysis (GRA) and Multiplicative Exponent Weighting (MEW) for vertical handover decision. In [3] author discuss that the vertical handover decision algorithm for heterogeneous wireless network, here the problem is formulated as Markov decision process. In [5] the vertical handover decision is formulated as fuzzy multiple attribute decision making (MADM).

In [8] their goal is to reduce the overload and the processing delay in the mobile terminal so they proposed novel vertical handover decision scheme to avoid the processing delay and power consumption. In [7] a vertical handover decision scheme DVHD uses the MADM method to avoid the processing delay. In [10] the paper is mainly used to decrease the processing delay and to make a trust handover decision in a heterogeneous wireless environment using T-DVHD.

In [11] a novel distributed vertical handover decision scheme using the SAW method with a distributed manner to avoid the drawbacks. In [14] the paper provides the four steps integrated strategy for MADM based network selection to solve the problem. All these proposals works mainly focused on the handover decision and calculate the handover decision criteria on the mobile terminal side and the discussed scheme are used to reduce the processing delay by the calculation process using MADM in a distributed manner. In our work we also did the MADM in D-VHD and T-DVHD schemes.

# 3. Vertical Handover Decision Schemes

Centralized vertical handover decision (C-VHD), Distributed vertical handover decision (D-VHD), Trusted Distributed vertical handover decision (T-DVHD) are the schemes used to reduce the processing delay between the mobile node and neighbour network while exchanging the information during the handover. In this paper, D-VHD and T-DVHD schemes are compared. MADM have several methods in literature [16]. TOPSIS is used in distributed manner for network selection.

## 3.1 Centralized vertical handover decision Schemes

In C-VHD, a Mobile Node (MN) exchanging the information message to the Neighbour networks mean processing delay was increased by distributing in centralized manner. When processing delay had increased overall handover delay increases. This is one of main disadvantage in C-DHD, so Distributed Vertical handover decision (D-VHD) schemes was proposed in [7][8].

## 3.2 Distributed vertical handover decision schemes

D-VHD is used to decrease the processing delay than the C-VHD schemes. This scheme handles the handover calculation to the Target visitor networks (TVNs). TVN is the network to which the mobile node may connect after the handover process was finished. In our work D-VHD takes into account : jitter, cost, bandwidth, delay as evaluation metrics to select a suitable VN which applied in TOPSIS method.

Network Selection Function (NSF):

The network selection decision process has denoted as MADM problem, NSF have used to evaluate from set of network using multiple criteria. The above mentioned parameters are used to calculate NSF. These parameters measure the Network Quality Value (NQV) of each TVN. The highest NQV value of TVN will be selected as Visited Network (VN) by the mobile node. The generic NSF is defined by using TOPSIS

(1)

Where, $NQV_i$ represents the quality of $i^{th}$ TVN. is the closeness to the ideal solution.

Based on the user service profile, handover decision parameters have assigns different "Weights" to determine the level of importance of each parameter. In equation (2), the sum of these weights must be equal to one.

(2)

Distributed Decision scheme:

The D-VHD is explained in the Fig. 3

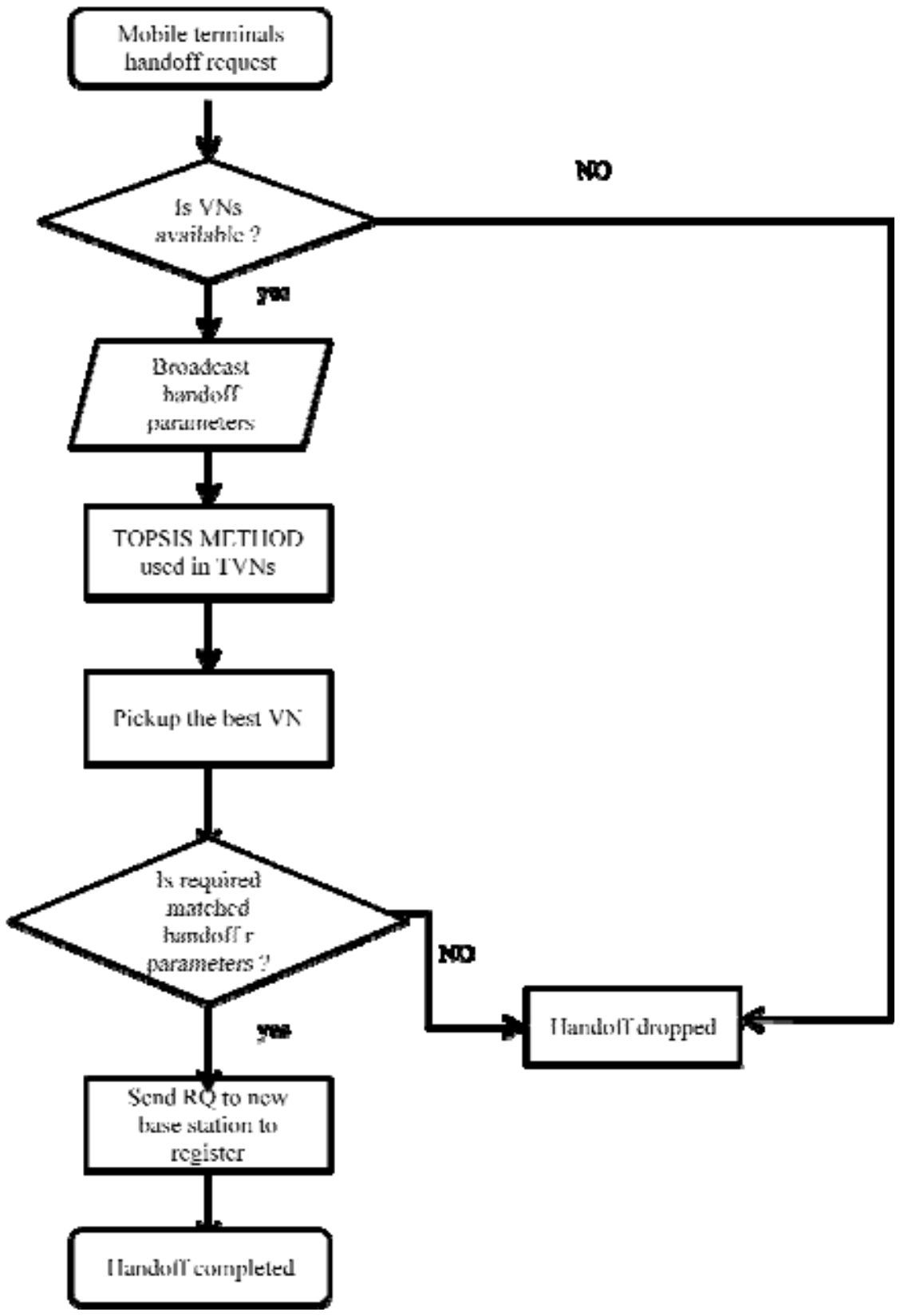

Fig.3 D-VHD Scheme

The handover decision metrics calculation is performed on the VNs, each VN applies the Topsis method using "(1)" on the required ($J_{req}$, $D_{req}$, $C_{req}$, $B_{req}$) and offered ($J_{off}$, $D_{off}$, $C_{off}$, $B_{req}$) parameters

## 3.3 Trusted Distributed Vertical Handover Decision schemes

Trusted handover decision and to avoid the unnecessary handover events are the important factors while exchanging the trusted information between networks and mobile node. The extension work of the DVHD scheme is T-DVHD scheme. The scheme is mainly introduced[10] for decreasing the processing delay than DVHD scheme.

The T-DVHD have the following steps:

Step1: Mobile terminal sends the handover request to available visitor networks.

Step2: Broadcast handover parameters includes the terminal request and handover matrix with the respective weight of the application

Step3: The handover decision phase calculation performed by VNs,

Step4: VN applies the TOPSIS method to find the NQV, with a set of parameters jitter, cost, bandwidth, delay.

Step5: normalized decision matrix was constructed by

$$r_{ij} = X_{ij}/\left(\sum X_{ij}^2\right) \quad (3)$$

for i=1,…,m; j=1,……,n

Step6: The weighted normalized decision matrix is constructed by

$$V_{ij} = w_j r_{ij} \quad (4)$$

Step7: Positive ideal and negative ideal solutions are determined by

Positive Ideal solution.

$$A^* = \{v_1^*, \dots\dots\dots, v_n^*\} \quad \text{where} \quad (5)$$

$$V_j^* = \{max_i(V_{ij}) if\ j \in J\, ; min_i(V_{ij}) if\ j \in J'$$

Negative ideal solution.

$$A' = \{v_1^*, \dots\dots\dots, v_n^*\}, \text{ where} \quad (6)$$

$$V_j^* = \{min_i(V_{ij}) if\ j \in J\, ; max(V_{ij}) if\ j \in J'$$

Step8: Separation measures for each alternative is calculated by

$$S_i^* = [\textstyle\sum_j (v_j^* - v_{ij})^2]^{1/2} \quad i = 1, \dots, m \quad (7)$$

$$S_i' = [\textstyle\sum_j (v_j' - v_{ij})^2]^{1/2} \quad i = 1, \dots, m \quad (8)$$

Step9: Relative closeness to the ideal solution $C_i^*$

$$C_i^* = S_i'/(S_i^* + S_i'), \quad 0 < C_i^* < 1$$

Step10: NSF is calculated for each TVNs by TOPSIS method and get the ‘best network’

Step11: Before sending request to connect a new base station trusted process is initiated

Step12: Level Of Trust (LOT) test function is tested to execute the handover

If LoTi >= threshold
    Connect to the TVNi
    Initiate Trust-test function
else if LoTi < threshold {
if (suitable-TVN available)
    i = i + 1
    test another network
else if (no suitable-TVN)
Handover blocked

Step13: The mobile Terminal executes with the proper TVN.

Step14: Trusted Test Function is initiated once the mobile terminal connects to the TVN

if Qoff < Qreq
$LOT_i$ = LOT – delta ;
else
$LOT_i$= $LOT_i$+ $delta^+$ ;

Step15: End of the trusted test function and T-DVHD.

## 4. Scenario of Vertical Handover

In this paper, our scenario was in Fig. 4, it explains that a cell coverage the area by WiMax technology and another cell coverage the area by WiFi and WiMax technology. A mobile terminal is overlapping with VoIP application between the cell coverage now mobile terminal intend to connect the appropriate visited network with the decision process.

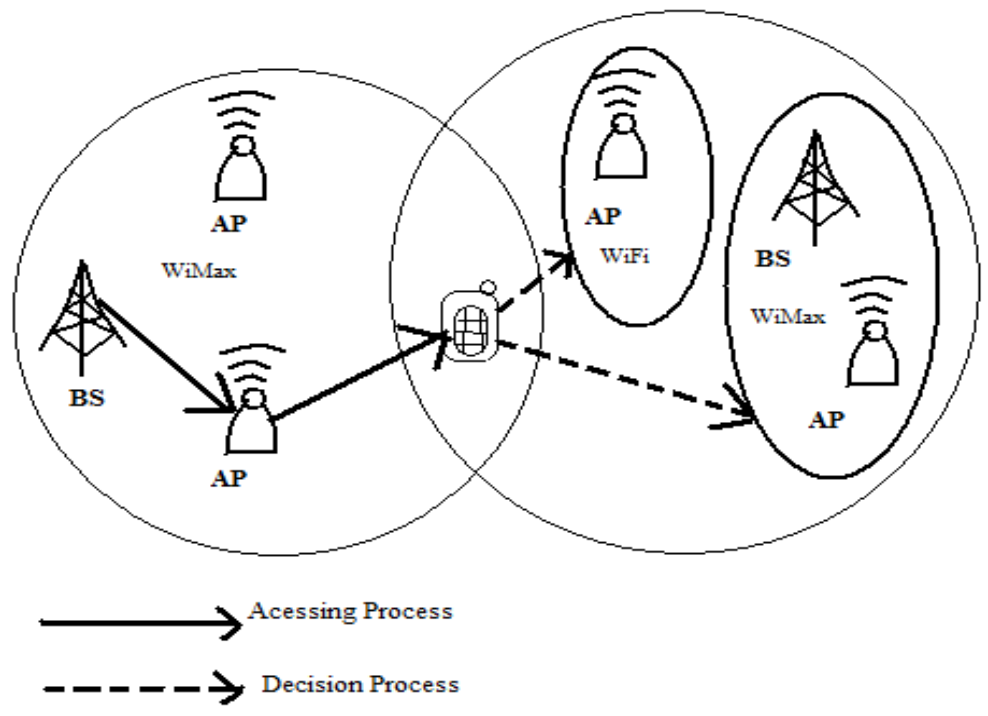

Fig. 4 Scenario of vertical handover

## 5. Simulation Result

In this section, the comparison of vertical handover decision scheme are compared and we provide the evaluation parameters used to analyze the performance T-DVHD schemes as well as the output of simulation. In our simulation we consider 7 mobile nodes are moving in an area covered by the heterogeneous wireless networks managed by 4 Base stations ($BS_i$=1,2,3,4). Mobility area covered by BS, supporting two types of technologies: WiMax and WiFi. These BS offer different characteristic in terms of coverage and QOS . VoIP is used as application in this simulation. Table 1 show the simulation metrics.

### 5.1 Evaluation parameters

There are different evaluations parameters are used, in order to evaluate our schemes. We have used:

- Processing Delay: It is a process which takes time by the terminal for making the decision towards which network to handover for network to handover
- Throughput: It is measured by the data are sent by the mobile node after a set of matching decision during a defined period.
- End to End Delay: It refers the time taken for a packet to be transmitted across a network from source to destination
- Handover Events: It reflects the number of handover achieved by the mobile terminal

TABLE I

TABLE FOR SIMULATION METRICS

| | |
|---|---|
| Topography | 500*500 |
| Mobile Node | 7 nodes |
| Base Station | 4 |
| CBR | .1 sec |
| Routing Protocol | DSDV |
| Packet Size | 1240bytes |
| Simulation time(s) | 200 sec |
| Wireless Standards | 802.16,802.11 |

### 5.2 Simulation Analysis

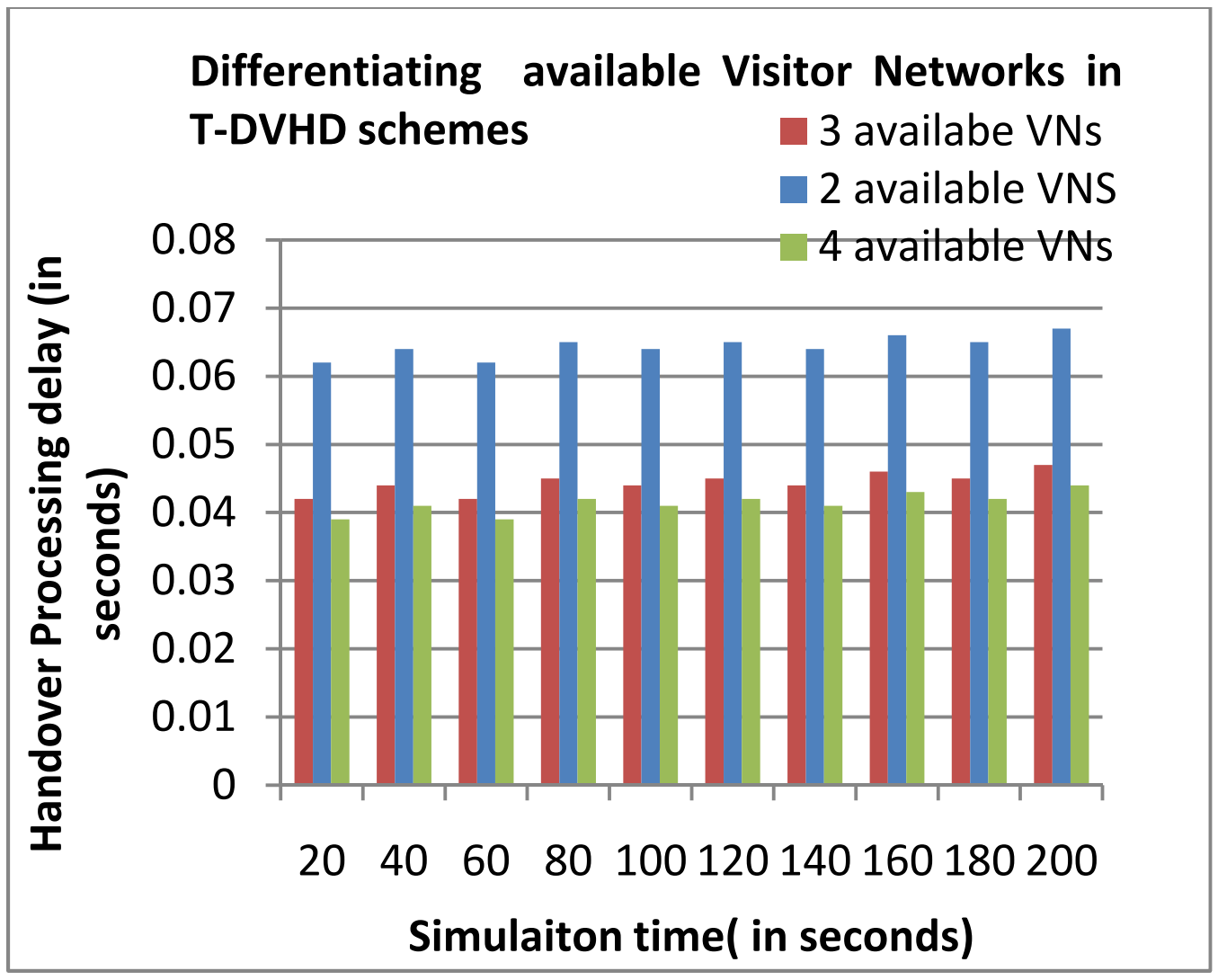

Fig.5 processing delay between available Visited Networks

Fig. 5 shows the Processing delay with available Visitor networks like 2, 3, 4 VNs by this we can analyze the time has taken for completing the whole handover process .

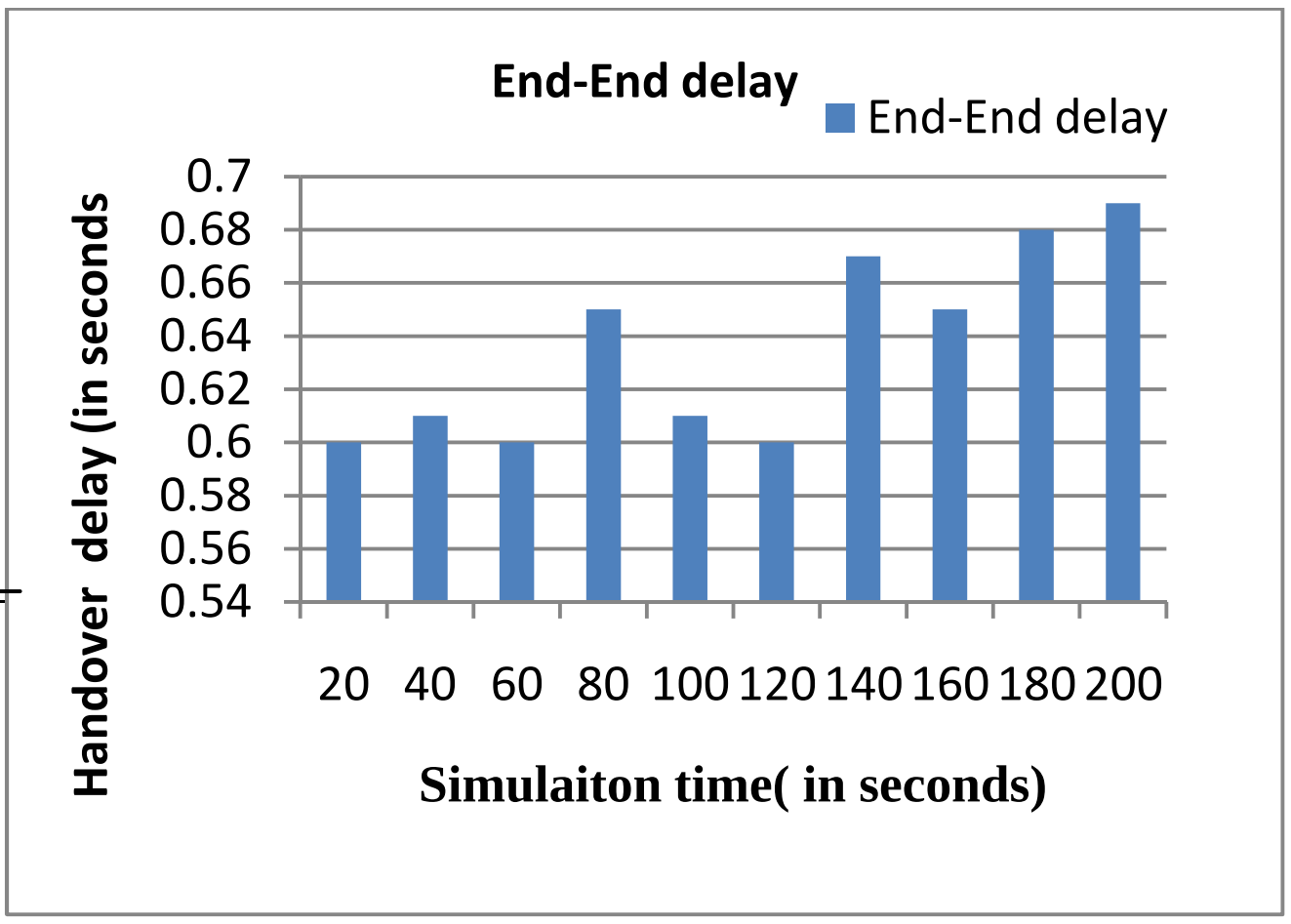

Fig. 6 End-End delay

End -End delay is sum of transmission delay, propagation delay and processing delay of number of links. End to End delay between the node and destination access point with required QOS service.

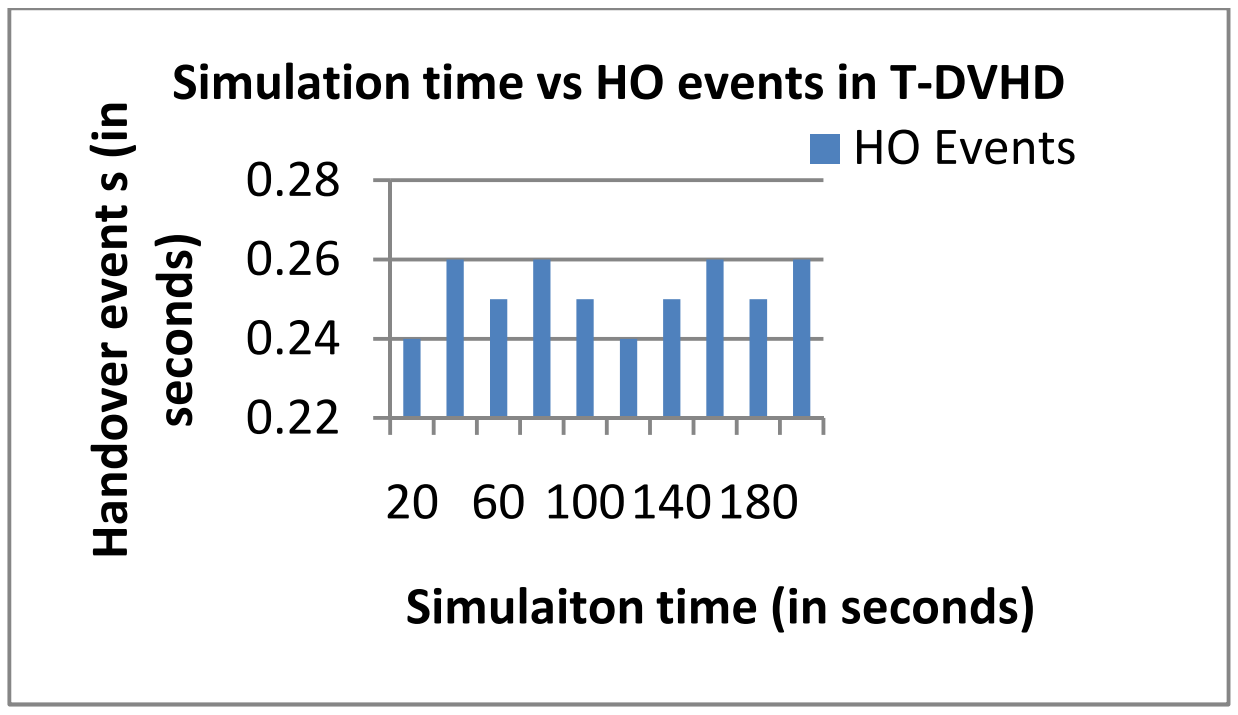

Fig. 7 Handover events

In fig. 7 Multiple handover events are occurred, when the mobile node chooses a TVN that provides falsified quality value (i.e. NQV). In case, another handover event may be performed as the switched VN doesn't provide the appropriate quality, which adds additional delay to the handover process it shows with the T-DVHD schemes.

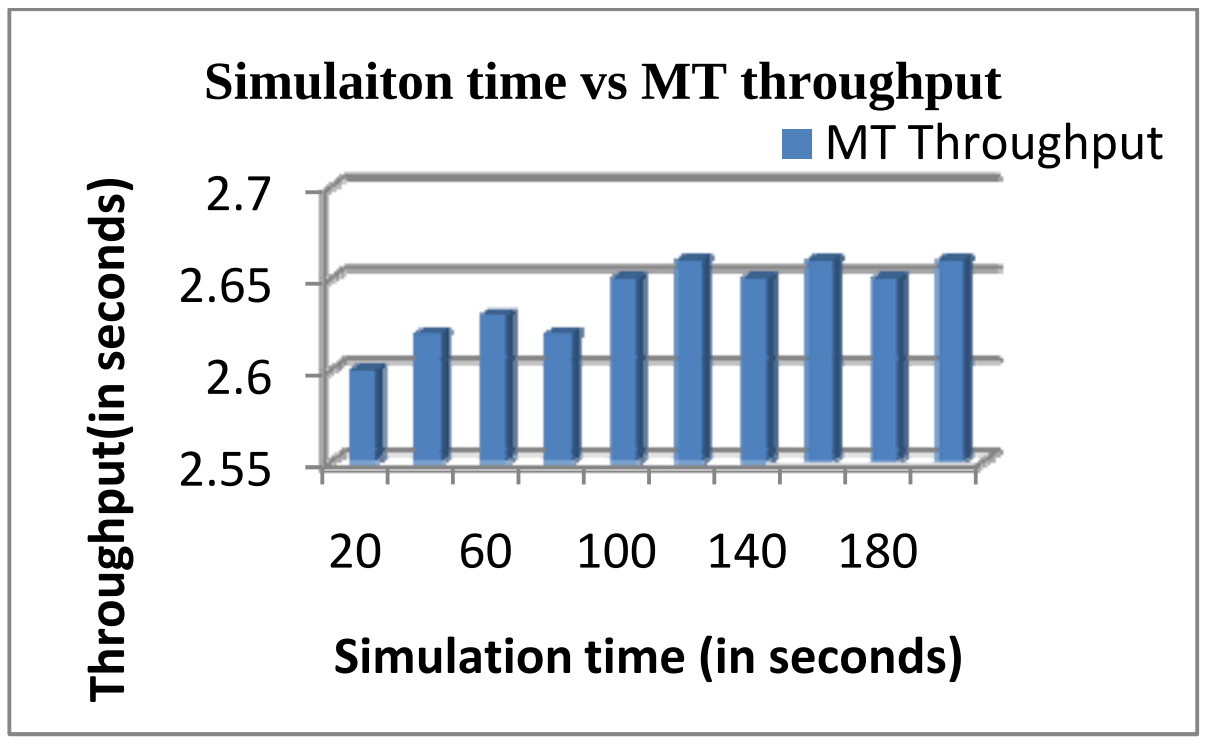

Fig. 8 Mobile Terminal Throughput

Throughput in Fig. 8 shows by the mobile terminal. Throughput is measured in bits per second. It calculated by Total Bytes Sent * 8 divide by Time Last Packet Sent - Time First Packet Sent here time is in seconds in T-DVHD schemes.

## 6. Conclusion

In our work, we have compared the schemes of vertical handover decision in the heterogeneous wireless networks. The main observation of the schemes to reduce the processing delays and a trust handover decision is done in a heterogeneous wireless networks. In this paper we proposed TOPSIS for the Vertical decision schemes for decision making to select the best network from the visitor network. Our main goal is in the decision phase of the handover phases to take decision to which VN the mobile terminal to connect by different decision algorithms.

**First Author** Mrs.K.Savitha received the Master of Science from the Bharathiar University, India in 2006, M.Phil Degree from periyar university, India in 2007. She is a Research Scholar in Department of computer science , Periyar University, Salem, India. She is pursing her Ph.D in Mobile computing. Her research area interest includes Networking, Multimedia.

**Second Author** Dr. C. Chandrasekar received his Ph.D. degree from Periyar University, Salem. He has been working as Associate Professor at Dept. of Computer Science, Periyar University, Salem – 636 011, Tamil Nadu, India. His research interest includes Wireless networking, Mobile computing, Computer Communication and Networks. He was a Research guide at various universities in India. He has been published more than 50 technical papers at various National/ International Conference and Journals.